\documentclass[runningheads]{svmult}

\usepackage{makeidx}   
\usepackage{graphicx}  
\usepackage{physprbb}  
\makeindex             

\begin{document}
\title*{Near-IR Spectroscopy and Population Synthesis\newline
of Super Star Clusters in NGC 1569}
\toctitle{Near-IR Spectroscopy and Population Synthesis
\protect\newline of Super Star Clusters in NGC 1569}
%
%
\titlerunning{NGC 1569 Super Star Clusters}
%
\author{Andrea M. Gilbert 
   \and James R. Graham }
\authorrunning{Andrea M. Gilbert \& James R. Graham}
\institute{Astronomy Department, University of California, Berkeley CA
94704, USA}

\maketitle              

\begin{abstract}
We present {\it H}- and {\it K}-band NIRSPEC spectroscopy of super
star clusters (SSCs) in the irregular starburst galaxy NGC~1569,
obtained at the Keck Observatory.  We fit these photospheric spectra
to NextGen model atmospheres to obtain effective spectral types of
clusters, and find that the information in both {\it H- and K}-band
spectra is necessary to remove degeneracy in the fits.  The light of
SSC B is unambiguously dominated by K0 supergiants (T$_{\rm eff}=4400
\pm 100$~K, $\log g=0.5 \pm 0.5$).  The double cluster SSC~A has
higher T$_{\rm eff}$ (G5) and less tightly constrained surface gravity
($\log g=1.3 \pm 1.3$), consistent with a mixed stellar population
dominated by blue Wolf-Rayet stars and red supergiants.  We predict
the time evolution of infrared spectra of SSCs using Starburst99
population synthesis models coupled with empirical stellar spectral
libraries (at solar metallicity).  The resulting model sequence allows
us to assign ages of 15--18~Myr for SSC~B and 18--21~Myr for SSC~A.

\end{abstract}

\section{Introduction}

At a distance of about 2~Mpc \cite{kt94}, the irregular galaxy
NGC~1569 is one of the nearest starbursts \cite{i88}.  It is rich in
ionized and neutral gas, and recently underwent a global starburst
that lasted at least 100~Myr and ended 5--10~Myr ago \cite{g98}.  The
starburst produced numerous H{\sc ii} regions and young star clusters,
and still drives an x-ray superwind \cite{w91,h95}.  One of the most
notable features of the starburst is the super star clusters (SSCs)
near the center of the galaxy \cite{as85,o94}. They are some of
the nearest and earliest-known examples of young, massive, compact
star clusters.  These SSCs, which may evolve into clusters resembling
present-day globulars \cite{m95}, have been found in all types
of starburst environments, from dwarf irregulars to galaxy mergers.

\section{Observations \& Data Reduction}

\begin{figure}[t]
\begin{center}
\includegraphics[width=.45\textwidth]{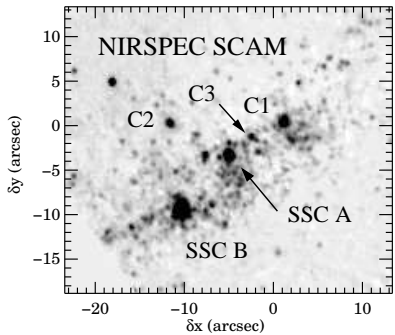}
\caption[gilbert_fig1.eps]{ NIRSPEC {\it K}-band image of NGC~1569 in
$0.''7$ seeing.  Clusters for which spectra were obtained are labeled}
\label{kimage}
\end{center}
\end{figure}

We obtained spectra and images of the SSCs in NGC~1569 on January 17,
2000 UT using the near-infrared spectrometer NIRSPEC on the Keck II
telescope at the W. M. Keck Observatory.  The night was photometric,
with seeing of about $0.''7$.  The {\it K}-band image (taken with the
N7 filter by NIRSPEC's slit-viewing camera) in Fig.~\ref{kimage} shows
the brightest central SSCs of which we obtained {\it K}-band spectra
(2.03--2.47~$\mu$m), along with many other clusters and stellar
sources.  For the brightest clusters, SSCs A and B, we also obtained
{\it H}-band spectra (1.49--1.78~$\mu$m).

\begin{figure}[t]
\begin{center}
\includegraphics[width=\textwidth]{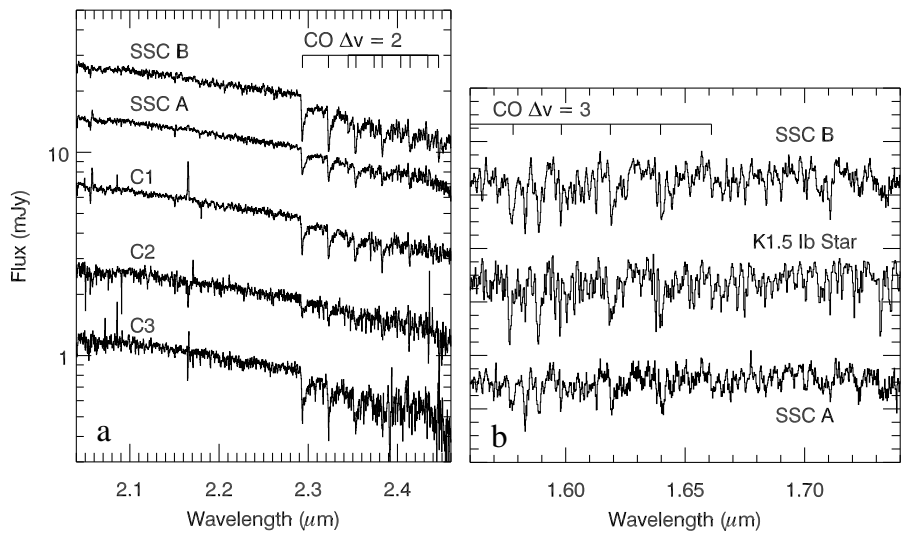}
\caption[gilbert_fig2.eps]{ NIRSPEC spectra of clusters in NGC~1569.
(\textbf{a}) {\it K}-band spectra.  Strong overtone CO bands signify
the presence of red supergiants.  Clusters A and C1 show weak
Br\protect$\gamma$ and He \protect{\sc{i}} emission, while C2 shows
both of these in absorption. (\textbf{b}) {\it H}-band spectra of SSCs
A and B with slopes removed, compared with the spectrum of a K1.5 Ib
supergiant \cite{m98}.  Note the correspondence of features between
star and cluster spectra, and the richness of {\it H} band}
\label{spectra} 
\end{center}
\end{figure}

Date were reduced in the usual fashion, and the optimally extracted
{\it K}-band spectra are shown in Fig.~\ref{spectra}a.  The strongest
features in {\it K} band are the CO first overtone bands longward of
2.295~$\mu$m.  These saturated bands are found in the cool
atmospheres of supergiants and giants.  Their presence implies that
the clusters contain stars that are at least 6 or 7~Myr old, since
that much time is required before the most massive stars evolve off of
the main sequence.

Stellar evolution models for a simple stellar population predict that
the hot blue stars capable of driving nebular emission will have
completely evolved off of the main sequence by the time red
supergiants appear.  Thus the conjunction of strong CO bands together
with weak Br$\gamma$ and He{\sc{i}} emission in the spectra of
Clusters A and C1 suggests that they are not instantaneous bursts.  In
the case of cluster C1 we believe that the weak emission lines are
explained by contamination from a nearby H{\sc ii} region \cite{w91}.
The weak nebular emission from SSC A (a double cluster \cite{dm97}) is
likely due to the Wolf-Rayet stars known to be present there
\cite{gd97}.  Thus the spectrum of SSC A cannot be explained with a
single simple stellar population according to current population
synthesis models.  Cluster C2, on the other hand, shows both
Br$\gamma$ and He{\sc i} in absorption, which may be due to a much
older population dominated by A stars and red giants.

Optimally extracted {\it H}-band spectra of SSCs A and B are shown in
Fig.~\ref{spectra}b, together with the spectrum of a K1.5 Ib
supergiant from the KPNO stellar atlases \cite{wh96,m98}.  A
comparison of the star and cluster spectra reveals that most of the
features in the cluster spectra are real, not noise, and that {\it
H}-band photospheric spectra are remarkably rich in strong metallic
and molecular features.  Only the CO bands stand out in the {\it
K}-band spectra, while many metal features in {\it H} band are as
strong as the second overtone CO bands.  Most of the features in {\it
H} band are blends of lines from CO, OH, and metals such as Fe, Si,
Al, and Mg.

\section{The Cluster Integrated Light}

The most massive, evolved members of a single-aged stellar population
tend to dominate its integrated light.  Thus the simplest approach to
interpreting the integrated light may be to ask, what is the dominant
spectral type of the cluster?  We attempt to characterize a cluster by
an effective spectral type by fitting its {\it H}- and {\it K}-band
spectra to a grid of NextGen model atmospheres \cite{h99}, in order to
determine an effective temperature, surface gravity, and metallicity
for each object.  The NextGen atmospheres were available for
spherically symmetric giant stars with a range in T$_{\rm eff}$ of
3000~K to 6800~K, a range in $\log{\rm g}$ of 0.0 to 3.5, and
metallicities of [Fe/H] $=$ 0.0 (solar), $-$0.3, $-$0.5, and $-$0.7.

\begin{figure}[t]
\begin{center}
\includegraphics[width=\textwidth]{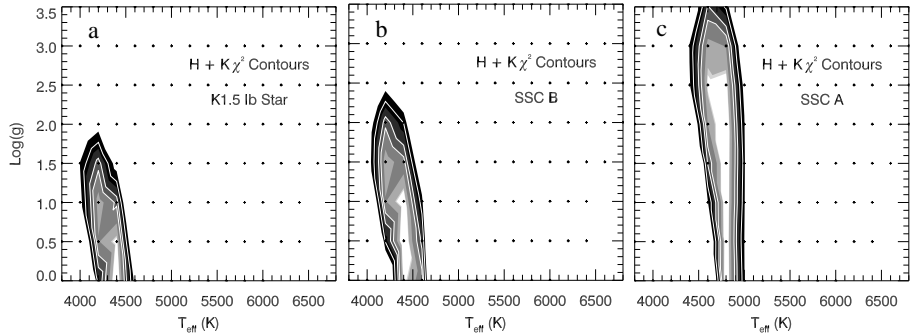}
\caption[gilbert_fig3.eps]{ \protect$\chi^2$ contours for fit of
NextGen models for: (\textbf{a}) spectrum of a K1.5 Ib star;
(\textbf{b}) SSC~B; (\textbf{c}) SSC~A.  White filled contour
indicates 1~sigma confidence level, and line contours represent 90,
95, and 99\% confidence limits.  Best-fit parameters for SSC B imply
an effective spectral type for the cluster light of K0 supergiant.
Best-fit parameters for SSC A imply an effective spectral type for the
cluster light of G5 supergiant with \protect$\log{\rm g} =1.3 \pm
1.3$.  The temperature and range of \protect$\log{\rm g}$ are
consistent with a mixed population of stars, either two short bursts
or an extended epoch of star formation}
\label{chi2} 
\end{center}
\end{figure}

We first test the utility of this procedure by fitting the model
atmospheres to an empirical spectrum of a star of known spectral type
to evaluate the precision and accuracy with which it selects the
atmospheric parameters.  Fig.~\ref{chi2}a shows the resulting $\chi^2$
contours for a fit to a solar-metallicity K1.5~Ib spectrum.  The
resulting parameters, T$_{\rm eff} = 4400 \pm 100$~K and $\log{\rm g}
= 0.0 \pm 0.5$, are a good match to those of the star.  Both {\it H}-
and {\it K}-band spectra were required in order to remove degeneracies
in the fits.

Next we fit the NextGen atmospheres at metallicity [Fe/H] $=-0.5$
(which is closest to that of NGC~1569, [Fe/H] $=-0.6$ to $-0.7$
\cite{gd97,ks97}) to the {\it H} + {\it K} spectra of the brightest
{\it K}-band cluster, SSC B.  The resulting $\chi^2$ contours in
Fig.~\ref{chi2}b have the same shape as those found for the star in
Fig.~\ref{chi2}a, indicating that the light of cluster B is heavily
dominated by stars with a very narrow range in spectral type.  The
effective spectral type for SSC B is that of a K0 supergiant, with
T$_{\rm eff} = 4400 \pm 100$~K and $\log{\rm g} = 0.5 \pm 0.5$.

For the double cluster, SSC A, the $\chi^2$ contours are less tightly
constrained in the fit parameters than for SSC B.  Fig.~\ref{chi2}c
shows them to be centered at a hotter T$_{\rm eff} = 4800 \pm 200$~K
and larger range of $\log{\rm g} = 1.3 \pm 1.3$, typical for stars of
types G5 I and G5 III.  It is unlikely, however, that such stars
dominate the cluster's emission, since optical/UV evidence indicates
that hot blue Wolf-Rayet stars are present together with the red
evolved stars creating the strong infrared CO bands.  Thus the inferred
effective spectral type determined from the stellar atmospheres may
simply result from the superposition of the two distinct populations.

\section{IR Spectral Population Synthesis}

Since a cluster consists of stars with a range of stellar masses,
luminosities, and temperatures, it is more informative to model the
integrated population directly than to focus on its effective spectral
type.  Thus we employ the technique of population synthesis to
calculate the distribution of stars in the H-R diagram of a cluster as
a function of time.  We then calculate the integrated spectrum of the
cluster by adding the appropriate numbers of stellar spectra -- either
empirical spectra or model atmospheres.

\subsection{Models}

All models in this paper were constructed using the updated
evolutionary synthesis code Starburst99 \cite{l99}.  The code
incorporates the most recent stellar evolutionary tracks from the
Geneva group at metallicities ranging from very metal-poor, to twice
solar metallicity \cite{s92,s93}, and it has been updated to allow the
use of isochrone synthesis.  Starburst99 is a particular set of
synthesis models which are optimized to reproduce properties of
galaxies with active star formation, so it puts most of the emphasis
on early evolutionary phases. Later phases, like AGB stars or white
dwarfs, are covered only crudely or not at all.

\begin{figure}[t]
\begin{center}
\includegraphics[width=\textwidth]{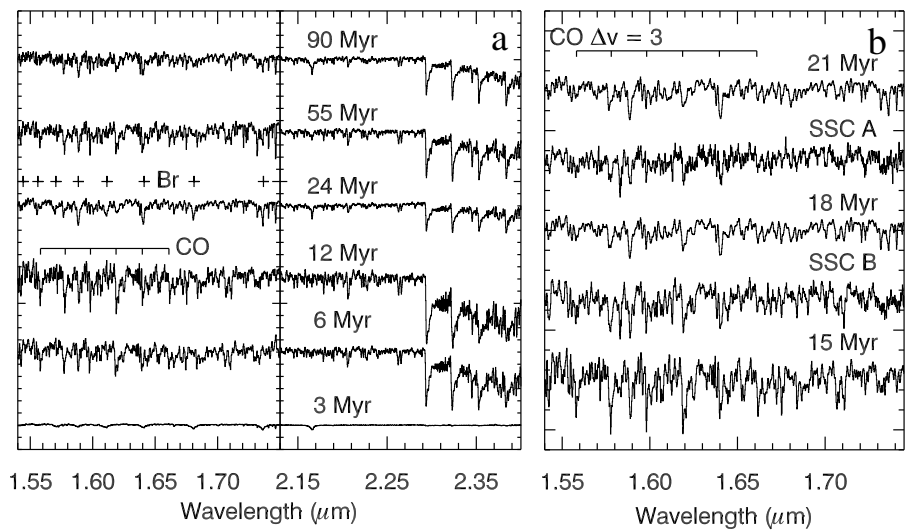}
\caption[gilbert_fig4.eps]{ (\textbf{a}) Sequence of model cluster
spectra as a function of time, calculated using Starburst99 and KNPO
stellar atlases.  (\textbf{b}) SSC A and B {\it H}-band spectra shown
in an age progression between synthetic cluster spectra (at solar
metallicity) at ages of 12, 15, and 18~Myr}
\label{mod} 
\end{center}
\end{figure}

Origlia et al. \cite{o99} show that low-metallicity tracks do not
reproduce the CO 1.62~$\mu$m and CO 2.29~$\mu$m indices of young LMC
clusters. However, if the fraction of time spent as a RSG during the
core-helium phase is forced to at least 50\%, and if the RSG
temperature is maintained to less than 4000~K, the models agree well
with the observations. Our modeling technique was modified according
to this prescription (Leitherer, private communication, 2000).

In order to generate model cluster spectra, we combine the Starburst99
models with the empirical libraries of stellar spectra obtained at
Kitt Peak by Wallace \& Hinkle \cite{wh96} and Meyer
et~al. \cite{m98}.  For a given cluster population, we add up the
spectra of component stars, and include nebular continuum emission
(but not the recombination lines) based on the number of ionizing
photons predicted for the cluster.  Thus we generate a time series of
model cluster spectra such as that shown in Fig.~\ref{mod}a for a
10$^6$~M$_{\odot}$ cluster with Salpeter IMF ranging from 0.1 to
100~M$_{\odot}$.  For the first few Myr, nebular emission powered by
the hottest stars dilutes the photospheric emission from the cluster,
but by an age of 6--7~Myr, the most massive stars have evolved off of
the main sequence to become red supergiants, whose spectra are marked
by deep CO bands.

Finally, we can place observed cluster spectra in an evolutionary
sequence by fitting them to the model sequences.  Figure~\ref{mod}b
displays the {\it H}-band spectra of SSCs A and B together with the
three model cluster spectra (15, 18, and 21 Myr) which most closely
resemble the observations.  Note the correspondence between features
and the decrease in their strength with time.  Since the models are
for solar metallicity clusters, they presumably have stronger metal
features at a given age than expected for a lower-metallicity cluster.
Thus the age estimates we derive from these models will be too large
during this epoch of the cluster's evolution.

\section{Conclusions}

We have presented new high-quality near-infrared spectra of several of
the SSCs in the nearby irregular starburst, NGC~1569, and demonstrated
the utility of the rich {\it H}-band spectral region for modeling
stellar populations.  We found that combining {\it H}- and {\it
K}-band spectra removed some of the degeneracy in fitting just one
band to model spectra.

We used population synthesis models together with model stellar
atmospheres and empirical stellar spectra to fit for the effective
spectral type of a cluster, and to generate sequences of synthetic
cluster spectra to help determine the ages of observed clusters.
Since the empirical libraries are only complete for solar metallicity,
we are constructing models that use model atmospheres at lower
metallicities that are more appropriate for systems like NGC~1569.

%

\end{document}